%% file: D4M_HPEC_2017.tex
\documentclass[conference]{IEEEtran}
\IEEEoverridecommandlockouts
\usepackage{cite}

\usepackage{graphicx}
\ifCLASSINFOpdf
\else
\fi
\usepackage[cmex10]{amsmath}
\usepackage{listings}
\usepackage{courier}
\lstnewenvironment{code}[1][]%
{
              \noindent
              \minipage{\linewidth}
              \medskip
              \lstset{frame=lrtb,captionpos=b,basicstyle=\ttfamily\footnotesize,breaklines=true,frame=single,#1}}
{\endminipage}

\usepackage{array}
\usepackage[caption=false,font=footnotesize]{subfig}
\usepackage{multirow}
\usepackage{textcomp}

\makeatletter
\def\footnoterule{\relax%
   \kern-5pt
   \hbox to \columnwidth{\hfill\vrule width 0.5\columnwidth height 0.4pt\hfill}
   \kern4.6pt}
\makeatother

\author{Lauren Milechin}

\begin{document}
%
\title{D4M 3.0: Extended Database and Language Capabilities}

\author{\IEEEauthorblockN{Lauren Milechin\IEEEauthorrefmark{1}, Vijay
    Gadepally\IEEEauthorrefmark{2}, Siddharth
    Samsi\IEEEauthorrefmark{2},  Jeremy Kepner\IEEEauthorrefmark{2},
    Alexander Chen\IEEEauthorrefmark{3}, Dylan Hutchison\IEEEauthorrefmark{4}}

\IEEEauthorblockA{\IEEEauthorrefmark{1}MIT EAPS, Cambridge, MA 02139
  \\ lauren.milechin@mit.edu}
\IEEEauthorblockA{\IEEEauthorrefmark{2}MIT Lincoln Laboratory, Lexington, MA 02420}
\IEEEauthorblockA{\IEEEauthorrefmark{3}MIT CSAIL, Cambridge, MA}
\IEEEauthorblockA{\IEEEauthorrefmark{4}University of Washington, Seattle, WA}
\thanks{This material is based upon work supported by the National Science Foundation under Grant No. DMS-1312831. Any opinions, findings, and conclusions or recommendations expressed in this material are those of the author(s) and do not necessarily reflect the views of the National Science Foundation.}
}


%


\maketitle

\input{abstract}


%
\IEEEpeerreviewmaketitle

\input{intro}

\input{d4m}

\input{sciDB}

\input{graphulo}

\input{julia}

\input{conclusion}

\input{acknowledgement}



%
\IEEEtriggeratref{19}
\bibliography{D4M_HPEC_2017}
\bibliographystyle{IEEEtran}

\end{document}

%% file: abstract.tex
\begin{abstract}
The D4M tool was developed to address many of today's data needs. This tool is used by hundreds of researchers to perform complex analytics on unstructured data. Over the past few years, the D4M toolbox has evolved to support connectivity with a variety of new database engines, including SciDB. D4M-Graphulo provides the ability to do graph analytics in the Apache Accumulo database. Finally, an implementation using the Julia programming language is also now available. In this article, we describe some of our latest additions to the D4M toolbox and our upcoming D4M 3.0 release. We show through benchmarking and scaling results that we can achieve fast SciDB ingest using the D4M-SciDB connector, that using Graphulo can enable graph algorithms on scales that can be memory limited, and that the Julia implementation of D4M achieves comparable performance or exceeds that of the existing MATLAB\textregistered{} implementation.
\end{abstract}

%% file: intro.tex
\section{Introduction}
\label{sec:intro}

There are many challenges to handling today's data, which hinder our ability to gain insight. Datasets are large, they are expanding quickly, and are not homogeneous. These challenges strain our systems and software. Tools are needed that can help store and index large data, support fast ingest, and can both manipulate and analyze varying types of data, whether it be numeric, text, or graph data. There are many databases which address one or more of these challenges. The D4M (Dynamic Distributed Dimensional Data Model) tool addresses these challenges by providing the ability to process heterogeneous incoming data, interaction with these databases through ingest and query, and the ability to analyze relevant data when needed.

The D4M tool is an analytical library for MATLAB\textregistered{}/GNU Octave that allows flexible data representation and manipulation \cite{d4m2012}. D4M's flexibility stems from the way it represents data: the mathematical structure of associative arrays.  Associative arrays can represent many different types of data, including graphical, numeric, and string data. Being a mathematical structure, associative arrays also support a variety of arithmetic and set operations that are facilitated through D4M and have a wide variety of uses \cite{accumuloEigensolver2015} \cite{cmd2015}. This combination of flexible data representation and mathematical operations yields something fairly powerful: associative arrays are amenable to performing linear algebraic operations on heterogeneous data. 

Another important property of D4M is its usefulness in the entire data analytics pipeline, including data access, ingest, and storage. In particular, D4M has been working seamlessly with the NoSQL database Accumulo for some time now, providing a simple means to bind to tables for ingesting and querying data as well as a schema for that data \cite{d4mschema2013}. Accumulo was built with fast ingest and query in mind, and past work has shown record-breaking ingest performance using the D4M's ingest and schema \cite{ingest2014}.

Numerous recent additions to D4M have prompted us to release the next version: D4M 3.0. The first two additions expand D4M's database connectivity and computation capabilities. This includes adding a connector to SciDB, a popular database for multidimensional scientific data, and integration with Graphulo, an extension on the Accumulo database that facilitates in-database graph analytics. The third addition is an implementation of D4M in the Julia programming language.

The advantage of adding these new features is flexibility. Data comes in so many different forms and sizes, and different algorithms and analytics stress different parts of the system and software. What could be a great solution on one dataset for one analytic might not work well for the same analytic on a different dataset. D4M provides a single interface to a large toolset, particularly with these new features. It can query data from the database that is the right fit, and it can enable fast, in memory computation as well as behind-the-scenes computation on a large dataset in a database, all in the language that the analyst is most comfortable with.

In the following sections of this work, we will describe these additions. Section \ref{sec:d4m} will introduce D4M more in depth. In Sections \ref{sec:sciDB} -- \ref{sec:julia} we will describe the technologies of SciDB, Graphulo, and Julia, and how these have been made accessible through or integrated with D4M, including a brief discussion of scaling results for each of these technologies.

%% file: d4m.tex
\section{D4M}
\label{sec:d4m}

D4M is open-source software that provides a convenient mathematical representation of the kinds of data that are routinely stored in spreadsheets and large key-value databases. Associations between multidimensional entities (tuples) using string keys and string values can be stored in data structures called associative arrays. For example, in two dimensions, a D4M associative array entry might be:

\vspace{6pt}
\centerline{\textbf{A}(\textquotesingle alice \textquotesingle, \textquotesingle bob \textquotesingle) = \textquotesingle cited \textquotesingle \hspace{4pt} or \hspace{4pt} \textbf{A}(\textquotesingle alice \textquotesingle, \textquotesingle bob \textquotesingle) = 47.0}
\vspace{6pt}

The above tuples have a 1-to-1 correspondence with their key-value store representations:

\vspace{6pt}
\centerline{(\textquotesingle alice \textquotesingle,\textquotesingle bob \textquotesingle,\textquotesingle cited \textquotesingle) \hspace{4pt} or \hspace{4pt}  (\textquotesingle alice \textquotesingle,\textquotesingle bob \textquotesingle,47.0)}
\vspace{6pt}

\begin{figure}[hh]
  \includegraphics[width=20pc]{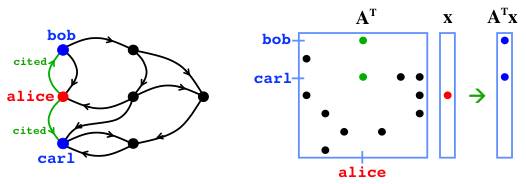}
  \caption{A graph describing the relationship between alice, bob, and carl (left). A sparse associative array A captures the same relationships (right). The fundamental operation of graphs is finding neighbors from a vertex (breadth-first search). The fundamental operation of linear algebra is matrix vector multiply. D4M associative arrays make these two operations identical.  Thus, algorithm developers can simultaneously use both graph theory and linear algebra to exploit complex data.}
  \label{fig:d4mfig}
\end{figure}

Associative arrays can represent complex relationships in either a sparse matrix or a graph structure (see Figure~\ref{fig:d4mfig}). Thus, associative arrays provide a natural data structure for performing both matrix and graph operations. Such algorithms are the foundation of many complex database operations across a wide range of fields~\cite{kepner2011}. Constructing complex composable query operations can be expressed by using simple array indexing of the associative array keys and values, which themselves return associative arrays:

\vspace{6pt}
\begin{tabular}{l l}
\centering
\textbf{A}(\textquotesingle alice \textquotesingle,:) & alice row \\
\textbf{A}(\textquotesingle alice bob \textquotesingle,:) & alice and bob rows \\
\textbf{A}(\textquotesingle al* \textquotesingle,:) & rows beginning with al \\
\textbf{A}(\textquotesingle alice : bob \textquotesingle,:) & rows alice to bob \\
\textbf{A}(1:2, :) & first two rows \\
\textbf{A} == 47.0 & subarray with values 47.0
\end{tabular}
\vspace{6pt}

The composability of associative arrays stems from their ability to define fundamental mathematical operations whose results are also associative arrays. Given two associative arrays A and B, the results of all the following operations will also be associative arrays: 

\vspace{6pt}
\begin{tabular}{c c c c c}
\textbf{A} + \textbf{B} \hspace{5pt}& \textbf{A} - \textbf{B} \hspace{5pt}& \textbf{A} \& \textbf{B} \hspace{5pt}& \textbf{A} $\mid$ \textbf{B} \hspace{5pt}& \textbf{A} * \textbf{B} \\
\end{tabular}
\vspace{6pt}

Measurements using D4M indicate these algorithms can be implemented with a tenfold decrease in coding effort when compared to standard approaches~\cite{d4m2012}.

%% file: sciDB.tex
\section{SciDB}
\label{sec:sciDB}

Part of D4M's capabilities include seamless interaction with a variety of database systems. Previous work largely focused on support of the scalable key-value store Apache Accumulo. With the recent trends of ``many sizes" and in-database analytics, we have extended the D4M connections to the relational databases PostGRES and MySQL and the NewSQL array data store SciDB.

SciDB is a database designed for multidimensional, scientific data. It uses an array data model and provides the ability to perform basic linear algebra operations on data within the database, without the need to query that data first \cite{scidbarch2011}. This is particularly useful when the data becomes too large to perform the operations locally in memory, or when querying the data or reading it from a file becomes too time consuming. It is also an ACID database, meaning it is suitable for applications with multiple users and a need for stricter database guarantees.

Because SciDB uses an array data model for storage, it is ideally suited for scientific data such as image, time series, weather, and sensor data. Data are stored in the user-defined coordinate system such that data close to each other in the coordinate system are stored in the same chunk on disk. This storage mechanism has a significant advantage for performing operations such as selecting ranges or joining multiple arrays. Additionally, SciDB can minimize number of files read because of the ability to specify overlaps in the chunks used to store data on disk. By appropriately specifying the array schema, it is possible to optimize the data access and query speeds in SciDB.

\begin{figure}[]
{\includegraphics[width=3in]{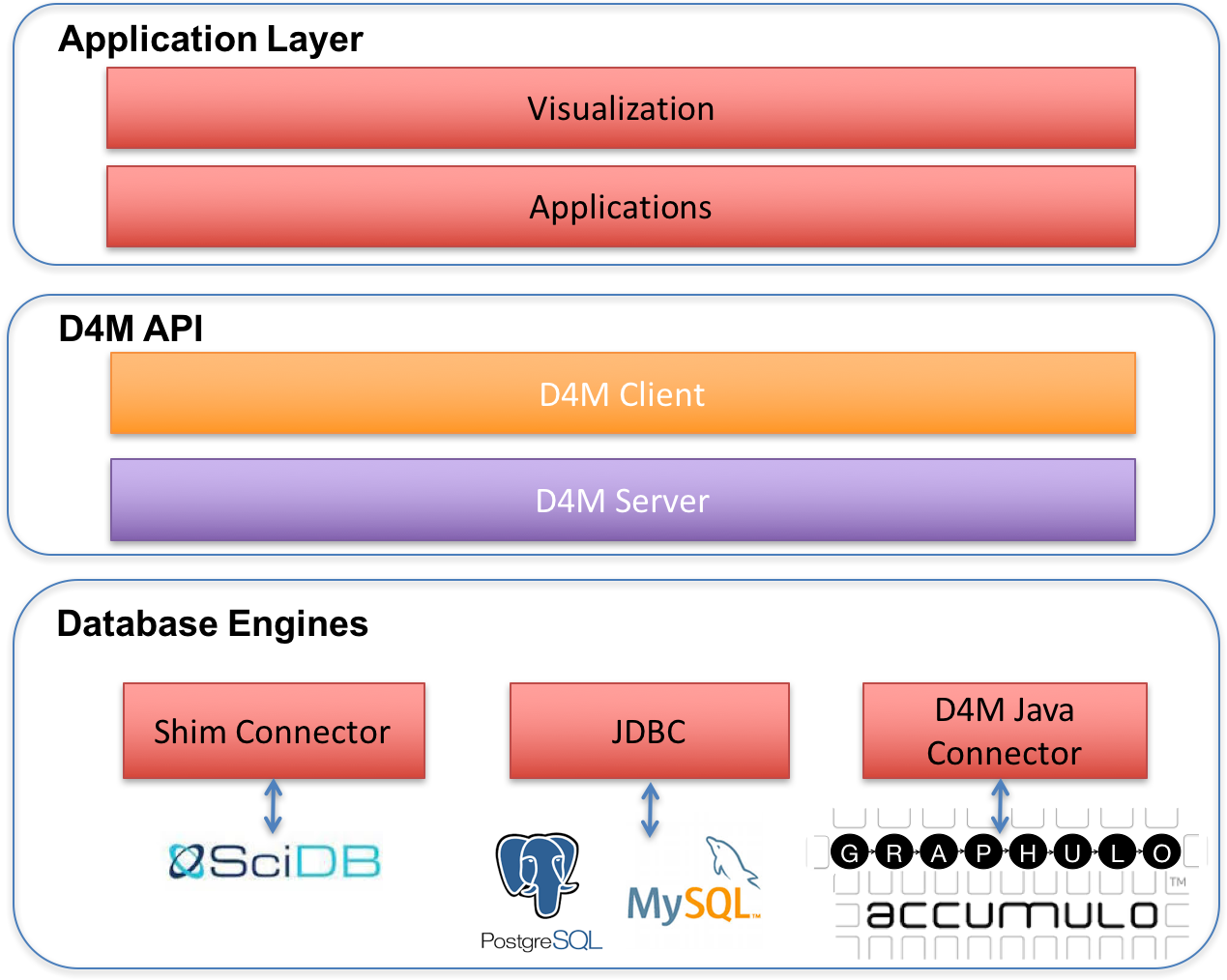}}
\centering
\caption{D4M architecture. D4M server bindings leverage various database connectors.}
\label{fig:d4mArch}
\end{figure}

The D4M-SciDB connector allows a user to connect to SciDB, bind to an array and ingest and query data, all using familiar associative array syntax. For the purpose of D4M, SciDB arrays are nothing but associative arrays and can be treated as such, much the same way Accumulo tables are treated as associative arrays in D4M. For example, we can ingest data using the same ``putTriple" command that is used to ingest data into Accumulo:

\begin{code}[frame=single, caption=Using D4M to ingest a 3D image into SciDB.]
DB = DBsetupSciDB('txg-testdb');
T = DB('vol3d<gray:uint8>row=1:4096,4096,0,col=1:4096,4096,0,slice=1:1000,1,0]');
im = imread('test-image.tif');
[nr, nc] = size(im);
[rowids, colids] = ind2sub([nr nc], [1:nr*nc]');
slicenum = 15*ones(size(ir));
T = putTriple(T, [rowids colids slicenum], im(:));
\end{code}

\noindent
and we can connect to that table containing the 3D image, and query that image for a specific sub-volume in three lines:

\begin{code}[frame=single, caption=D4M query to SciDB to extract a sub-volume.]
DB = DBsetupSciDB('txg-testdb');
T = DB('vol3d<gray:uint8>row=1:4096,4096,0,col=1:4096,4096,0,slice=1:1000,1,0]');
v = T(100:300, 100:500, 10:100);
\end{code}

 The D4M-SciDB connector has demonstrated fast ingest into SciDB \cite{scidbd4m2016}. This work explored parallel ingest performance for large volumetric image data on both single and two-node SciDB instances. Peak ingest performance was found to be nearly 3 million inserts per second.

%% file: graphulo.tex
\section{Graphulo}
\label{sec:graphulo}

\begin{figure*}[!t]
{\includegraphics[width=3.45in]{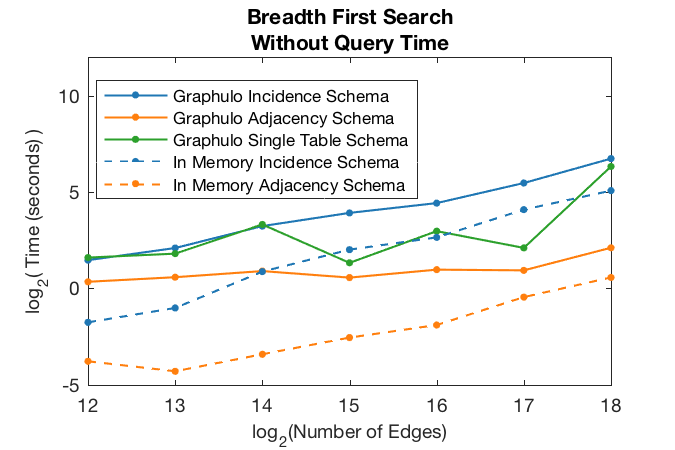}
\label{fig:bfs-woQ}}
{\includegraphics[width=3.45in]{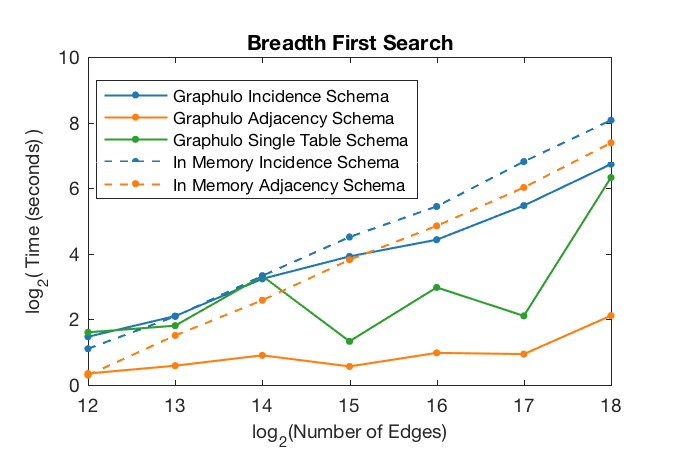}
\label{fig:bfs-wQ}} \\

{\includegraphics[width=3.45in]{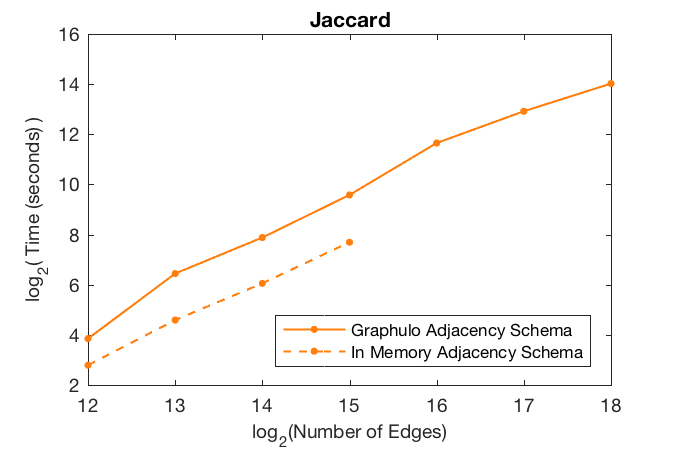}
\label{fig:jaccard}}
{\includegraphics[width=3.45in]{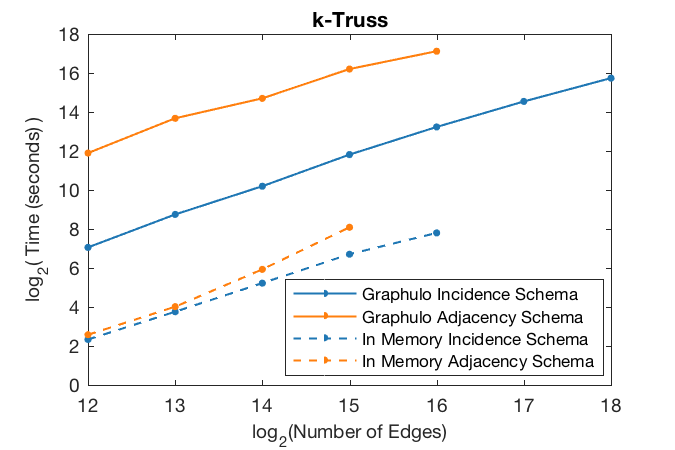}
\label{fig:ktruss}} \\
\caption{Timing for running three Graphulo implementations of graph algorithms through D4M compared to in memory (on a laptop) runtime, where possible. Each is run for all available schemas. The first row compares degree-filtered breadth first search, where the second plot includes the time taken to query for the graph from Accumulo. In the second row the first shows scaling results for calculating a matrix of Jaccard Coefficients, and the final plot shows results for the k-Truss algorthm.}
\label{fig:graphuloResults}
\end{figure*}

The second addition to the D4M toolbox is the ability to perform fast in-database analytics in the NoSQL key-value store Apache Accumulo. As part of the D4M 3.0 release, Graphulo, an extension on Accumulo, is also included. Graphulo provides in-database sparse linear algebra operations outlined in the GraphBLAS, implemented as Accumulo server-side iterators. These operations can be combined to implement a variety of graph algorithms.

The D4M-Graphulo tool allows users to describe their graph analytics in the familiar GraphBLAS constructs and leverage the parallel infrastructure of Apache Accumulo to perform these operations directly in database without first transferring a partial set of results to local memory. These operations, such as matrix multiply, enable many useful graph algorithms, including breadth first search (BFS), Jaccard coefficient, and k-Truss~\cite{graphalgs2015}. While Graphulo is written in Java, these operations and graph algorithms can be called from D4M.

Graphs are often represented in a variety of different schemas, depending on the graph, application, and user preference. Graphulo supports three common schemas for graph representations. The first and most simple is the adjacency matrix representation, where each row and column index represents a vertex, or node, of the graph, and the entries within the matrix represent the existence of an edge or its weight, if the graph is weighted. In Graphulo, this adjacency matrix would make up one table, with an additional table containing vertex degrees. This additional table is not only useful for query planning, but it is used in the implementation of graph algorithms, such as degree-filtered BFS. The second schema is based on the incidence matrix representation and is the same as the D4M Schema \cite{d4mschema2013}. In the incidence matrix, each row represents an edge, and each column represents a vertex. The values within the matrix indicate which vertices participate in, or are incident to, each edge. Additional tables are also included in this schema. The first is the transpose of the incidence matrix, to enable fast search of vertices, as Accumulo searches most efficiently by row key. This schema, like the Adjacency schema, also includes a degree table. The final schema is known as the Single Table schema and consists a single table containing two types of entries: degree entries that hold the degrees of each vertex and edge entries that indicate which edges exist by listing the pairs of vertices that participate in that edge. Each schema has its advantages, so Graphulo aims to support all three where possible.

To use Graphulo through D4M, the first step is to bind to a database, requesting a Graphulo object:

\begin{code}[frame=single, caption=Setting up Graphulo in D4M.]
[DB,G] = DBsetupLLGrid('graphulo-db');  
\end{code}

\noindent Once the Graphulo object is available, a number of graph operations and algorithms are available. For example, BFS, Jaccard, and k-Truss can each be called using D4M as follows:

\begin{code}[frame=single, caption={Graphulo call to BFS, Jaccard, and k-Truss in D4M.}]
G.AdjBFS(Tadj, v0, k, Rtable, RtableT, TadjDeg, degColumn, degInColQ, minDegree, maxDegree);
G.Jaccard(Aorig, ADeg, Rfinal, filterRowCol, Aauth, RNewVis);
G.kTrussAdj(Aorig, Rfinal, k, filterRowCol, forceDelete, Aauth, RNewVis);
\end{code}

Graphulo have been shown to scale well to multi-node Accumulo instances \cite{graphulobench2016} and outperform the client-side alternative in many cases. Our extensive performance results indicate that D4M-Graphulo can be used in cases where data size makes operations impossible to complete client-side due to memory constraints \cite{graphuloMM2015} \cite{newsqlgraphulo2016}. We have performed significant experiments with the D4M-Graphulo tool and have compared it to numerous parallel processing paradigms. These performance results focus on matrix multiply, which is a core GraphBLAS operation.

In Figure \ref{fig:graphuloResults}, we show some timing results for the graph algorithms BFS, Jaccard, and k-Truss using different schemas, where available. Degree-filtered BFS was performed on five randomly chosen vertices, with a minimum degree of 1 and maximum degree of 100, and k-Truss was run with a $k$ of 3. Graphs were generated using the Graph500 unpermuted power law graph generator \cite{d4mpowerlaw2015} with scale ($s$) 12-18 and an average degree ($d$) of 16, producing graphs with $2^s$ vertices and $d*2^s$ edges. With the exception of the first plot in Figure \ref{fig:graphuloResults}, each plot includes the time it took to query for the graph for each local computation. The Graphulo operations were initiated through D4M locally and performed on a single node Accumulo instance running on the MIT SuperCloud dynamic database system~\cite{prout2015enabling}, and the operations identified as ``Local" were performed in MATLAB\textregistered{} using D4M implementations on a standard laptop with 16 GB of RAM.

Local BFS outperformed Graphulo's BFS with five initial vertices. However, when the time to query the graph from Accumulo is included, Graphulo surpasses local performance after Scale 15. Local operations for computing Jaccard and performing k-Truss also out performed Graphulo, but they ran out of memory after scale 15, 16 for kTruss using the Incidence schema. Because these algorithms take longer to perform than BFS, including query time did not make as big a difference in runtime. Note that numbers are not provided for the Graphulo k-Truss execution on the Adjacency schema above scale 16. Scale 17 ran for several days before the operation was terminated.

%% file: julia.tex
\section{Julia}
\label{sec:julia}

D4M can be implemented in any language that provides support for sparse linear algebra operations. Our previous 
versions have been implemented in MATLAB\textregistered{} and required either MATLAB\textregistered{} or GNU Octave to use. A recent addition to the D4M tool chain is the \texttt{D4M.jl} package which allows users to use D4M in the Julia programming language.

Julia is a newer language developed for both high performance and high-level dynamic programming \cite{juliaFresh2014}. Typically, languages are either high performance, low level, but difficult for development, or high-level and easy for development, but without the performance that a low-level language usually provides. The Julia developers aimed to create a language that is both high-level and high performance. Julia contains both state of the art numeric computation libraries and a state of the art Just-in-Time (JIT) compiler built on Low Level Virtual Machine (LLVM) \cite{juliaLang2012}. In this way, Julia's low-level functionality can be optimized, so that the user will not resort to using other languages as low-level building blocks. By leveraging the selected chain of modern programming language technologies within Julia, the Julia community has been rapidly expanding the high-level functions of Julia without compromising in performance. Julia has been shown to be effective in high performance computing \cite{juliaHPC2016}. It is for these reasons that Julia is a good candidate for a D4M implementation.

\texttt{D4M.jl} takes the approach of providing the functionality of D4M, but with familiar Julia syntax and conventions to the Julia programmer, much the same way D4M-Matlab is designed to be intuitive to the MATLAB\textregistered{} user. This makes \texttt{D4M.jl} more useable to the Julia community. For example, because Julia uses square brackets for indexing rather than the parentheses that are used in MATLAB\textregistered{}, \texttt{D4M.jl} follows this convention. It also provides the ability to convert DataFrames, a popular representation of tabular data, to Associative Arrays and vice versa:

\begin{code}[frame=single, caption=Converting between DataFrames and Associative Arrays in D4M.jl.]
DataFrame df = DataFrame(A::Assoc)
Assoc A = Assoc(df::DataFrame)
\end{code}

\noindent DataFrames are a common way data is represented in Julia, and so this translation allows \texttt{D4M.jl} to be easily integrated into current projects without the need to write a parser from scratch.

Benchmarking work has shown \texttt{D4M.jl} operations to have comparable performance to MATLAB\textregistered{} D4M operations, and in some cases surpass the D4M-Matlab implementation \cite{julia2016}. This work tested four D4M operations: three types of matrix multiply (traditional and versions where the column keys or values are concatenated, rather than removed or multiplied and summed, respectively) and matrix addition, executed on both a laptop and the MIT SuperCloud environment. These operations were performed on increasing matrix sizes, and showed that in all cases \texttt{D4M.jl} performance was either comparable or exceeding D4M-Matlab performance. The \texttt{D4M.jl} toolbox is open sourced and available for download \cite{d4mJuliaDownload2016}.

%% file: conclusion.tex
 \section{Conclusions and Future Work}
The upcoming D4M 3.0 release incorporates many developments over the past three years. The D4M-SciDB connector allows access, ingest, and querying of the array datastore SciDB. Integration of Graphulo enables many graph algorithms that were not possible due to memory constraints by executing these algorithms within the Apache Accumulo database. Finally, the new \texttt{D4M.jl} brings D4M to the Julia community by providing a D4M implementation in the Julia language. We believe that greater support for database engines, support for in-database operations, and wider language support are important additions to the D4M system. 

There are many avenues for future work. First, we are working with the Intel Science and Technology Center for Big Data on integrating D4M operations within the BigDAWG polystore system~\cite{gadepally2016bigdawg}. Developing such integration, will allow D4M users to leverage a much wider set of database systems and operations. Another area for future work is developing connections with sparse matrix accelerators such as the Graph Processor~\cite{song2013novel}. We believe that such integration can help with within-core operations that can complement the out-of-core Graphulo system. We are also working to develop a D4M-Graphulo connector for Apache Pig, a platform for analyzing large datasets. This connector will use D4M constructs to allow users to connect to Accumulo from Pig and initiate Graphulo commands from that platform. Finally, \texttt{D4M.jl} does not currently have the same database connectivity as the MATLAB\textregistered{} implementation, and we are looking to add this capability.

%% file: acknowledgement.tex
\section*{Acknowledgment}

The authors acknowledge the following individuals for their contributions: Michael Stonebraker, Sam Madden, Bill Howe, David Maier, Chris Hill, Alan Edelman, Charles Leiserson, Dave Martinez, Sterling Foster, Paul Burkhardt, Victor Roytburd, Bill Arcand, Bill Bergeron, David Bestor, Chansup Byun, Mike Houle, Matt Hubbell, Mike Jones, Anna Klein, Pete Michaleas, Julie Mullen, Andy Prout, Tony Rosa, and Chuck Yee. 